# Taxes in a simple wealth distribution model by inelastically scattering particles

Sebastian D. Guala

Instituto de Ciencias, Universidad Nacional de General Sarmiento,

Gutierrez 1150, Los Polvorines (B1613GSX), Buenos Aires, Argentina

sguala@ungs.edu.ar

**Abstract**

In this work we use an inelastic scattering process of particles to propose a model able to reproduce the salient features of the wealth distribution in an economy by including taxes to each trading process and redistributing that collected among the population according to a given criterion. Additionally, we show that different optimal levels of taxes may exist depending on the redistribution criterion.

## 1 Introduction

More than a century ago Pareto [1] proposed a law of distribution of individual wealth in a society. It states that the tail of the wealth distribution follows a power-law $P(w) \approx w^{-(1+\nu)}$. Here, $P(w)$ is the amount of people possessing wealth $w$. The exponent $\nu$ is the Pareto exponent and generally takes values between 1 and 3 [2, 4, 5]. It is also known that for low and medium income, $P(w)$ decays exponentially or in a log-normal way [2, 3, 4]. Since the same behavior was found for income distribution in a society [6, 7], in the following and for the sake of concreteness, we will refer to "wealth", but "income" should be equally applicable.

As a consequence of the analysis of wealth and income distributions in different societies from available real data [6], it is well-known now that the distribution of the richest 5% of the population has a power-law tail [2, 4, 5], while the majority (around 95%) low-income distribution fits well to Gibbs or log-normal form [8].

There has been several attempts to model a simple economy [9, 10, 11, 12, 13, 14, 15, 16, 17, 18, 19, 20, 21, 22, 23], which involve a wealth exchange process that produce a distribution of wealth similar to that observed in real economies. Most of works are particularly interested in microscopic models of markets where the economic activity is considered as a scattering process. Special attention have received those models centered on savings in the trading process [6, 7, 8] which reproduce features of real wealth distributions.

This kind of models is analogously thought as either elastically or inelastically scattering particles. Inelastic scattering of particles was studied in the context of granular materials [24, 25, 26] by the inelastic variant of the Maxwell model [27, 28, 29, 30]. The conclusion of these studies is that a self-similar solution of the kinetic equations exist, which is not stationary in time at individual level, but the system converges to time-independent parameters. The model introduced here studies how inelastic binary collisions may be assumed as the application of taxes and its redistribution can reproduce the salient features of empirical distributions of wealth without regard to, in principle, any another consideration. Based on the mentioned saving models and from the majority log-normal distribution point of view, we propose a simple granular closed-system model in which the collisions are inelastic and the loss of energy is eventually redistributed among the particles of the system according to certain "state" criterion.

A parallel aim of this work is to evaluate the response of the simple model when the scope of feasible rules is extended to include taxes, specially related to income distribution, according to realistic observations.

## 2 The model

Imagine a closed economic system with total amount of money (wealth) $W$ and total number $N$ of agents, both constant. It is, neither production nor death/birth of agents occurs. The only economic activity is confined to trading. Agent $i$, possess a wealth $w_i(t)$ at time $t$. Time changes after each trading. Any trading involves two steps: first, two randomly chosen agents $i$ and $j$ exchange their money in "inelastic" way such that a fraction of their exchanged wealth is lost by taxes (Figure 1):

$$w_i(t + \frac{1}{2}) = (1-f)\epsilon_{ij}(w_i(t) + w_j(t))$$
$$w_j(t + \frac{1}{2}) = (1-f)(1-\epsilon_{ij})(w_i(t) + w_j(t))$$
$$w_i(t + \frac{1}{2}) + w_j(t + \frac{1}{2}) = (1-f)(w_i(t) + w_j(t))$$
$$w_k(t + \frac{1}{2}) = w_k(t) \text{ for } k \neq i, j$$

where $w_i(t), w_j(t), w_k(t) \geq 0$ for all $i, j, k$ and $t$, $f$ is the fraction of trading lost by taxes and $\epsilon_{i,j}$ is a random fraction ($0 \leq \epsilon_{i,j} \leq 1$). Secondly, taxes are redistributed among individuals according to certain rule (i.e., equally distributed among the population, distributed among the poorest fraction, etc.):

$$w_r(t+1) = w_r(t + \frac{1}{2}) + \begin{cases} \frac{f}{|S|}(w_i(t) + w_j(t)), & \forall r \in S \\ 0, & \text{otherwise} \end{cases}$$

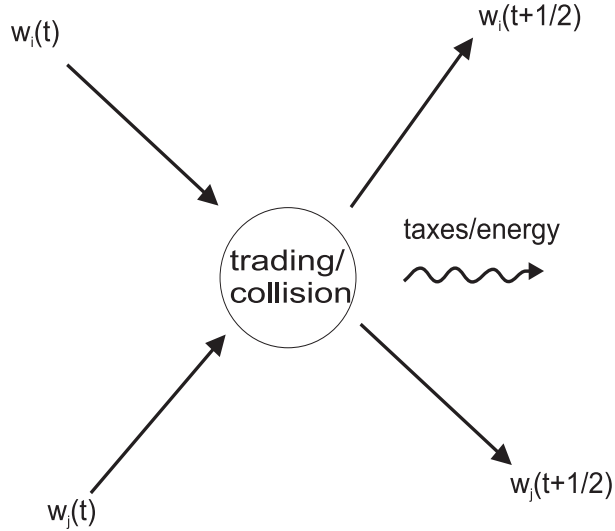

Figure 1: Schematic picture of the first step of the scattering process. Agents $i$ and $j$ exchange their wealth in the economy and pay taxes for their exchanges, which are redistributed among the population.

for $S$ being the subset of population benefited by the redistribution policy; conserving the total wealth $W = \sum_i^N w_i(t)$. From now on, we will use $P(w)$ normalized as a probability distribution function of wealth $w$.

As we can see here, at $f = 0$ the steady-state wealth distribution of the economy becomes a Gibbs' one as reported in [31], with modal value $w_m = 0$. For $S = N$ and $f \neq 0$ (i.e., when all the agents are equally benefited by the redistribution), the equilibrium distribution becomes asymmetric unimodal with $w_m$ of $P(w)$ shifting away from $w_m = 0$, reaching a maximum $w_m$ and moving back to $w_m = 0$ as $f \longrightarrow 1$, as shown in Figure 2.a. In this organization of the economy induced by taxes with a global egalitarian perspective is very significant the way in which the fraction of paupers decreases until certain level of taxes and most people end up close to the average wealth in the economy.

In this sense, slopes of the log-normal cumulative probability $Q(w)$, defined as $Q(w) = \int_w^\infty P(w')dw'$, for the after-mode part $P(w)$ in Figure 2.b decrease as taxes increase, attaining a minimum and then increasing with taxes tending to behave as to $f = 0$. The modal value of wealth distribution seems to behave on a contrary way, according to Figure 2.a. Both figures may suggest an optimal level of taxes, in which the distribution is relatively more egalitarian. Figures 3.a and 3.b show modal wealths $w_m$ of $P(w)$ and slopes of log-normal $Q(w)$ as a function of $f$, respectively. Besides, we consider the case in which the money collected by taxes is selectively redistributed. Figures 3.c and 3.d show the modal values $w_m$ of $P(w)$ and $P(w)$, respectively, when that money is uniformly redistributed among the subset $S$ conformed by the 20% poorest of the population. Note from Figure 3.c that $w_m \geq 1$ for $f \geq 0.3$. This fact suggests that a huge majority belong to the richest portion of the population and a few individuals belong to the poorest one, shown in Figure 3.d. As a consequence, the wealth of the multitudinarious richest portion is slightly grater than the average wealth $W/N = 1$, getting closer to a socialist distribution.

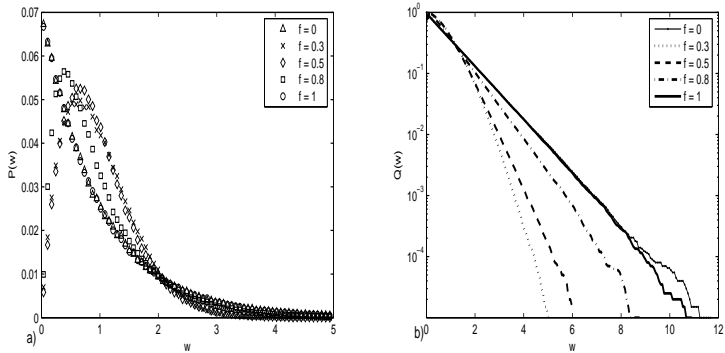

Figure 2: a) Wealth distribution $P(w)$ and b) cumulative probability $Q(w)$ for different taxes $f$; $N = 1000$ and average wealth $W/N = 1$.

## 3 Conclusions

We proposed a new rule for the simple binary trading model in the search of adding further complexities, which might appear in real economies. Besides, by this work we evaluated the robustness of that model for allowing other potentially-realistic rules, holding the same qualitative behavior. In this case, a granular model was analyzed in which inelastic binary collisions were consider in a closed system where the after-collision lost energy is captured by the molecules in it with the aim of emulating taxes paid in each trade operation and then redistributed by the corresponding State. Observations have been made about the wealth distribution obtained fitted as a log-normal one. The optimum level of taxes $f \cong 0.325$ shows a more egalitarian economy for a model with non-selective redistribution. In the case where the redistribution is focused on the poorest individuals we observed a significant majority with wealth around the average one.

Further research in this sense should be centered in an analytical treatment of the model and in more complex instances which include simultaneously taxes into the well-known models of saving propensity, binary trading according to the possibilities of the poorest agent of both, evasion propensity according to level of taxes, etc., in order to get models from a comprising conception.

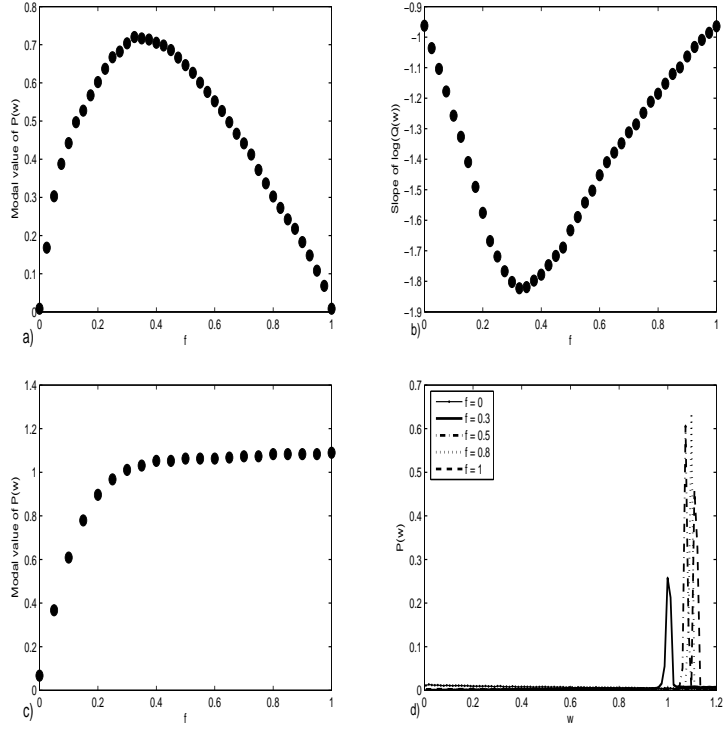

Figure 3: a) Modal values of wealth distribution $P(w)$ and b) slopes of log-normal curves of $Q(w)$ as a function of $f$. c) Modal values of wealth distribution $P(w)$ when taxes are redistributed among the 20% poorest of the population and d) $P(w)$ as a function of $f$; $N = 1000$, and average wealth $W/N = 1$.